\documentclass[prl,aps,amssymb,amsmath,showpacs,nofootinbib]{article}
\usepackage{amssymb,amsmath}
\begin{document}
\date{January 17, 2003}
\title{{\bf Equality of the  Inertial and the Gravitational Masses 
for a Quantum Particle}}
\author{Jaros{\l}aw Wawrzycki\footnote{Electronic address: 
Jaroslaw.Wawrzycki@ifj.edu.pl
or jwaw@th.if.uj.edu.pl}
\\Institute of Nuclear Physics, ul. Radzikowskiego 152, 
\\31-342 Krak\'ow, Poland}
\maketitle
\newcommand{\ud}{\mathrm{d}}

\begin{abstract}

We investigate the interaction of the gravitational field with a 
quantum particle. We derive the wave equation in the curved 
Galilean space-time from the very broad Quantum Mechanical
assumptions and from covariance under the Milne group. 
The inertial and gravitational masses are equal
in that equation.  So, we give the proof of the equality for 
the non-relativistic quantum particle, without applying the 
equivalence principle to the Schr\"odinger equation and without 
imposing any relation to the classical equations of motion. 
This result constitutes a substantial strengthening 
of the previous result obtained by Herdegen and the author.

\vspace{1ex}

\footnotesize

   
\end{abstract}

\section{Introduction}

We assume that the coefficients of the Schr\"odinger
equation describing a particle in the fundamental fields  are local functions 
of the space-time coordinates; compare the scalar
or the electromagnetic potential for example. At the the same time
we expect that the coefficients of the wave equation of a freely falling
quantum particle in the gravitational field are local functions of space-time 
coordinates which can be built up in a local way from the fields
which describe the space-time. In this paper we consider only the 
non-relativistic case and represent the Newtonian gravity in the 
geometric way compatible with the equivalence principle \cite{Car}, \cite{Tra},
\cite{Dau}. Now the rather unexpected fact comes. Namely, if we assume
the wave equation to be generally covariant, then the inertial mass of 
the quantum particle has to be equal to its  gravitational mass.
Strictly speaking the covariance under the Milne group is sufficient.  
Note that the equality is proved independently of the equivalence principle 
applied to the wave equation. Moreover, as we will see, the form of the wave 
equation is almost uniquely determined by the covariance condition. 
The covariance condition should be distinguished
from the symmetry condition. The covariance with respect to a group means that
the transform of a solution to the wave equation is a solution to
the transformed wave equation. The symmetry under the group means
in addition that the group is the symmetry group of the absolute 
elements of the theory in question, compare \cite{And} for the standard 
terminology. The geometrical objects describing the space-time structure
are the absolute objects in our case on account of the fact that we neglect
the influence of the particle on the space-time. As we know,
this is justified up to the second order effects even for the 
electromagnetic interactions, compare the semi-classical theory of radiation.
The gravitational interaction is extremely weak.
By this the neglecting
the influence on the space-time is justified.     

It should be stressed here that the equality of the inertial 
and the gravitational mass would not hold if the standard covariance 
condition with respect to the Galilean group were considered. 
In general it is impossible to restrict the covariance 
group to the Galilean one if the gravity is present. But in the wide
class of space-times, those which correspond to isolated systems,
one can restrict the covariance group in a consistent
manner by distinguishing the class of asymptotically
inertial frames.            
 
In our previous paper \cite{AHJW} we get the same result, but with 
the help of the equivalence principle. That is we have assumed that the
gravitational effects may be ``transformed away'' by an appropriate
reference frame. By this the ordinary ``flat'' form can be given
to the wave equation by an appropriate coordinate transformation.
Here we do not make any use of the equivalence principle.

\section{Derivation of the wave equation}

Our assumptions are, more precisely, as follows: (i) The quantum particle,
when its kinetic energy is small in comparison to its rest energy $mc^{2}$, 
does not exert any influence on the space-time structure. (ii) The Born 
interpretation for the wave function is valid, and the transition 
probabilities in the Newton-Cartan space-time which describes geometrically 
Newtonian gravity, are equal to the ordinary integral over a simultaneity
hyperplane and are preserved under the coordinate transformations. (iii) The 
wave equation is linear, of second order at most, generally covariant, and
can be built in a local way with the help of the geometrical objects 
describing the space-time structure. In fact the conditions
(i), (ii), (iii) are somewhat interrelated.
For example the linearity of the wave equation is deeply connected with
the Born interpretation. It will be shown below, that the equality of inertial 
and gravitational masses for a spin-less non-relativistic 
particle is a consequence of (i), (ii), (iii).    

We derive the most general form of the wave equation fulfilling (i), (ii),
(iii). A great simplification follows from the fact that in the 
Newton-Cartan theory the absolute elements exist\footnote{The simplification 
does not exist in the  General Relativity. But we expect the same result 
for the bare masses in the relativistic theory. 
But the argumentation should be within the path-integral 
formalism for the Feynman propagator of a structureless particle, 
see \cite{Bievre}.}.
The absolute elements fix the privileged, \emph{i.e} non-rotating Cartesian, 
coordinates. In those coordinates the absolute elements take on a particularly
simple form. The transformations connecting any two privileged coordinate 
frames form a group called the Milne group \cite{Milne}. 
The existence of such privileged frames largely simplifies the investigation 
of the consequences of general covariance. The simplification has its source 
in the fact that the absolute elements are invariant under the Milne group 
and have the same \emph{canonical} form in all privileged frames. 
This implies that  the Newtonian 
potential $\phi$ is the only object, which describes the geometry and does
not trivially simplify to a constant equal to 0 or 1, in these coordinates.
The wave equation written in the privileged coordinates is covariant under 
the Milne group in consequence of the general covariance. 
The Milne group is sufficiently rich to determine the wave 
equation as the covariant equation under the group if we use the 
assumptions (i), (ii), (iii). We confine ourselves then, to the privileged 
frames and the Milne group of transformations $r$:
\begin{equation}\label{milne}
(t,x_{j}) \to (t +b, R_{j}^{i}x_{i} + A_{j}(t)), 
\end{equation}
where $R_{j}^{i}$ is a rotation matrix,
and $A_{j}(t)$ are ``arbitrary'' functions of time. Strictly speaking
it is sufficient to consider a finite-dimensional subgroup of
the Milne group of polynomial $A_{j}(t)$ of appropriately high degree. 
From the space-time geometry and the Born interpretation it follows 
that the transformation law for the wave function $\psi(X)$  of a spin-less 
particle has the general form
\begin{equation}\label{wavetra}
\psi'(X)= T_{r}\psi(X)  = e^{-i\theta(r,X)}\psi(r^{-1}X),
\end{equation}    
where we denote all the space-time coordinates by $X$, and the
Milne transformation by $r$. $\theta(r,X)$ is a real function. 
The exponent $\xi(r,s,t) = \theta(rs,X) - \theta(r,X) - \theta(s, r^{-1}X)$ 
in the relation
\begin{equation}\label{xi}
T_{r}T_{s} = e^{i\xi(r,s,t)}T_{rs},
\end{equation}
depends on $r,s$ and also on the time $t$. The nontrivial time dependence of 
$\xi$ originates from the gauge freedom of the wave equation which cannot 
\emph{a priori} be excluded if the gravitational field is present.
A well-known Bargmann's theory \cite{Bar} provides a classification
of exponents $\xi$ which are time independent. In Ref. \cite{JW} a general
classification of $\xi$'s has been presented for a time dependent
exponents as is  necessary for the present article. In fact this explicit
time dependence of the exponent $\xi$ is necessary to account for the
experiments \cite{COW}, \cite{Kasevich}, \cite{Nesvi}. That is,
the gauge freedom is needed, compare \cite{DK}, \cite{Kuch}
or \cite{Waw} for the simplest wave equation in the gravitational field 
which does possess the gauge freedom.
The exponent $\xi$ could not depend on the time if the Milne group 
was a symmetry group of the wave equation. The Milne group is not a symmetry 
group of the wave equation because the Newtonian potential does not possess 
any symmetry in general.  
The classification of all possible $\xi$-s gives us the 
classification of all possible $\theta$-s in (\ref{wavetra}). 
The most general $\theta$ has the form \cite{JW} 
\begin{equation}\label{theta}
\theta(r,X) = - \gamma_{1}\frac{d}{dt}A_{j}x^{j} - \ldots - 
\gamma_{n}\frac{d^{n}}{dt^{n}}A_{j}x^{j} + \tilde{\theta}(r,t), 
\end{equation}
where $\tilde{\theta}$ is any function of $r$ and time $t$ and
$\gamma_{i}$ in (\ref{theta}) are some arbitrary constants. 
The coefficients $a, b^{i}, \ldots$
in the wave equation 
\begin{displaymath}
\Big[ a\partial_{t}^{2} + b^{i}\partial_{i}\partial_{t} + 
c^{ij}\partial_{i}\partial_{j} + f^{i}\partial_{i} + 
d\partial_{t} + g \Big]\psi  = 0,
\end{displaymath}
are local functions of the potential and cannot depend on the 
arbitrary high order derivatives of the potential. From (iii) it follows then, 
that the coefficients are functions of the potential and its derivatives up to 
a (say) $k$-th order. 
In the mathematical terminology this means that $a, \ldots, g$ are 
\emph{differential concomitants} of the potential, see \cite{Acz}.  
We assume in addition that $k=2$. We do not lose any generality 
by this assumption, beside this the whole reasoning could be applied 
for any finite $k$. But the case with $k>2$ would not be physically
interesting. Namely, it is \emph{a priori} possible that the derivatives 
of second order are discontinuous, such that the derivatives of order $k>2$ 
do not exist, at least the classical geometry does allow such a
situation. On the other hand there does not exist any mathematical 
obstruction for a discontinuity of the wave
equation coefficients, take for example the wave equation with 
the "step-like" potential. Then the assumptions
about the existence of higher oder derivatives which are not necessary 
for the space-time geometry, confines
our reasoning rather then generalizes it. 
Now, we insert the formulas (\ref{wavetra}) and (\ref{theta}) to the 
covariance condition of the wave equation. The covariance condition gives 
us the transformation formulas for the coefficients in the wave equation under 
the Milne group. For the coefficients $a, b^{i}, c^{i j}$ the transformations 
reads
\begin{equation}\label{trb}
{b'}^{i}(X')=R^i_j b^{j}(X)+2a(X) \dot{A}^{i},
\end{equation}
\begin{equation}\label{a}
a'(X')=a(X),
\end{equation}
\begin{equation}\label{trc}
{c'}^{ij}(X')=R^i_s R^j_k c^{sk}(X)+a(X) \dot{A}^{i} 
\dot{A}^{j}+b^{k}R^i_k \dot{A}^{j},
\end{equation}
where the dot stands for the time derivative.
 The formula (\ref{trb}) is valid in each privileged 
system and for any potential, and implicitly at any space-time point. 
Let us take then, such a system and let $X_{o}$ be any (but fixed) 
space-time point.
We consider the formula (\ref{trb}) for the special transformations
with $R={\bf 1}$ and $b=0$, $\vec{A}(t)=A(t)\vec{n}$, where  $\vec{n}$ 
is a constant in space and time space-like unit vector.
The analysis of (\ref{trb}) for $A(t)=\lambda (t-t_{o})^4$, then for 
$A(t)=\lambda (t-t_{o})^3$ and at last for $A(t)=\lambda (t-t_{o})^2$
with any value of the parameter $\lambda$
gives the following general form for the coefficient $b^{k}$
\begin{displaymath}
b^{k}(X) = b^{k}(\phi, \partial_{i}\phi, \partial_{i}\partial_{j}\phi, 
\partial_{t}\phi, \partial_{j}\partial_{t}\phi, \partial_{t}^{2}\phi)
= b^{k}(\partial_{i}\partial_{j}\phi).
\end{displaymath} 
It means, that $b^{k}$ is a vector concomitant, at least under rotations, 
spatial inversion and spatial reflections, 
of a tensor $\partial_{i}\partial_{j}\phi$ of valence 2. As is well known 
from the theory of geometric objects, such a vector concomitant 
has to be zero. The argumentation is as follows. Take any 
privileged system and any point $X_{o}$. Apply now the space 
inversion with the origin in $X_{o}$,
\emph{i.e.} $R=-{\bf 1}$ and $ \vec{A}=2\vec{x}_{o}, \, b=0$. 
Then, (\ref{trb}) at $X_{o}$ with this inversion gives 
the equation: $\vec{b}(X_{o})=-\vec{b}(X_{o})$ because the valence 
of $\partial_{i}\partial_{j}\phi$ is even 
and $\partial_{i}\partial_{j}\phi$ does not change the sign under 
the inversion. Because the point $X_{o}$ 
and the privileged reference frame can be chosen in an arbitrary 
way the concomitant $\vec{b} = 0$. 
From (\ref{trb}) immediately follows, that also $a=0$.
We have reduced our equation to the following form
\begin{displaymath}
[c^{ij}\partial_{i}\partial_{j}+d\partial_{t}+f^{i}\partial_{i}+g]\psi=0.
\end{displaymath}
Covariance condition of the equation under the Milne group gives
the following transformation law for $f^{j}$
\begin{equation}\label{22}
{f'}^{j}(X')=R^{j}_{i}f^{i}(X) - d\dot{A}^{j}-2ic^{ij}\partial_{i}\theta.
\end{equation}
First of all let us take notice of the fact that $\gamma_{j}=0$ for $j>4$. 
Indeed, let $X_{o}=
(\vec{x_{o}},t_{o})$ be any point. We apply now a Milne transformation  
for which all derivatives of 
$\vec{A}(t)$ disappear at $t_{o}$ with the exception of the $j$-th order 
derivative. For example, we can choose
such a transformation as in the preceding considerations $A(t)=(t-t_{o})^{j}$.
Then, we insert the transformation to the
law (\ref{22}). Because the derivatives of the order higher then the 4-th 
do not appear in the transformation laws for
$\phi, \partial_{i}\phi, \ldots, \partial_{t}^{2}\phi$, then (\ref{22}) 
at $X_{o}$ implies that $\gamma_{j}=0$.  
Note, that $f^{i}$ is an algebraic function of the potential 
and its finite order derivatives with the order less or equal 
then $k = 2$. The natural number $n$ in (\ref{theta})  is then finite 
and it is equal $k+2 = 4$ at most. We define the following object
\begin{displaymath}
\widetilde{f}^{j} \equiv f^{j}  
+2i\gamma_{2}c^{ij}\partial_{i}\phi + 
2i\gamma_{3}c^{ij}\partial_{t}\partial_{i}\phi,
\end{displaymath}
with the following transformation law 
\begin{displaymath}
{\widetilde{f}}^{\, ' i}(X')=R^{i}_{s}\widetilde{f}^{s}(X)
-(d -2i\gamma_{1}c^{sj}R^{i}_{s} -
\end{displaymath} 
\begin{equation}\label{trf'}
-2i\gamma_{3}{R^{-1}}^{i}_{s}{R^{-1}}^{q}_{p}c^{sk}
\partial_{q}\partial_{k}\phi\delta^{pj})\dot{A}_{j} 
-2i\gamma_{4}c^{sj}R^{i}_{s}\ddddot{A}_{j}.
\end{equation} 
A similar analysis as this applied to $b^{k}$ shows that
$\widetilde{f}^{k}=0$, or equivalently
\begin{displaymath}
f^{k}=-2i\gamma_{2}c^{ij}\partial_{j}\phi
-2i\gamma_{3}c^{ij}\partial_{j}\partial_{t}\phi.
\end{displaymath}
But this is possible only if $\gamma_{2}=\gamma_{3}=0$ 
or equivalently, only if $f^{k}=0$. Indeed,
applying the transformation laws for $\partial_{i}\phi$ and 
$\partial_{i}\partial_{t}\phi$ to the above 
formula one gets the transformation law for $f^{k}$
\begin{displaymath}
{f'}^{i}(X')=R^{i}_{s}f^{s}(X)
-2i\gamma_{3}{R^{-1}}^{i}_{s}{R^{-1}}^{q}_{p}
c^{sk}\partial_{k}\partial_{q}\phi\dot{A}^{p}-
\end{displaymath}
\begin{equation}\label{22'}
-2i\gamma_{2}R^{i}_{s}c^{sk}\ddot{A}_{k}-
2i\gamma_{3}R^{i}_{s}c^{sk}\dddot{A}_{k}.
\end{equation}
Consider the Milne transformation  
with $R \ne {\bf 1}$ and $A^{i}(t)=(t-t_{o})^{2}n^{i}$ such that 
$v^{j}\equiv c_{o}^{ij}n_{i} \ne 0$, where $c_{o}^{ij} = c_{ij}(X_{o})$. 
This is possible because 
$c_{o}^{ij} \ne 0$.  Note, that if 
$c_{o}^{ij}=0$, the analysis for $f^{k}$ reduces to the case such as with 
$b^{k}$ and $f^{k}=0$.  
Comparing (\ref{22'}) with (\ref{22}) at $X_{o}$ for this 
Milne transformation one gets
\begin{displaymath}
\gamma_{2}R^{j}_{i}v^{i}=\gamma_{2}v^{j},
\end{displaymath}
for all orthogonal $R$ and $\vec{v} \ne 0$, which means that $\gamma_{2}=0$. 
In the similar way, but with
$A^{i}=(t-t_{o})^{3}n^{i}$, one shows that $\gamma_{3}=0$. 
Summing up $f^{k}=0$.
Now, looking back to the transformation law (\ref{22}) we realize that 
\begin{equation}\label{dertheta}
\partial_{j}\theta=-\gamma_{1}\dot{A}_{j}, c^{ij}=c\delta^{ij}, 
c \equiv \frac{d}{2i\gamma_{1}},
\end{equation}
where $c$ is a scalar field: $c'(X')=c(X)$ which follows from the fact that 
$c^{ij}$ is a tensor field, compare (\ref{trc}) and recall that $b^{k}=0$ 
as well as $a=0$. Note that $\gamma_{1}$ is the inertial
mass of the particle in question and by this $\gamma_{1} \ne 0$. 
The wave equation must be of the form 
\begin{displaymath}
\Big[\frac{k}{2\gamma_{1}}\delta^{ij}\partial_{i}\partial_{j}+ik\partial_{t}
+g \Big]\psi=0,
\end{displaymath}
where we introduce $ik \equiv d$.
The covariance condition of the equation gives the following transformation 
law of $g$
\begin{displaymath}
g'(X')=g(X)-\frac{k\gamma_{1}}{2}\dot{A}_{i}\dot{A}^{i}
+k\gamma_{1}\dot{A}_{i}\dot{A}^{i}-
\end{displaymath}
\begin{displaymath}
-k\partial_{t}\widetilde{\theta}(A^{k},t)-k\gamma_{1}\ddot{A}_{i}x^{i}.
\end{displaymath}

Let us define a new object 
\begin{displaymath}
\Lambda(X)=g(X)+\gamma_{1}k(X)\phi(X).
\end{displaymath}
It is clear that the transformation law of $\Lambda$ is as follows
\begin{displaymath}
\Lambda'(X') = \Lambda(X) + 
\frac{k\gamma_{1}}{2} \dot{A}_{i}\dot{A}^{i} 
- k\partial_{t}\widetilde{\theta}(A^{k}, t).
\end{displaymath}
Both $\widetilde{\theta}$ and $\Lambda$ taken separately are not uniquely 
defined. This is because the potential $\phi$ is determined 
up to a time dependent additive term, namely, the gauge freedom term. 
So, one can assume any form for $\widetilde{\theta}$ by an appropriate gauge 
redefinition of $\phi$ changing the first one by $G(t)$ and the second one by
$(1/\gamma_{1})\dot{G}(t)$. 
Assume then, that $\widetilde{\theta}$ is chosen in such a way that 
$\partial_{t}\widetilde{\theta}=\gamma_{1}/2\dot{A}_{i}\dot{A}^{i}$. 
After this the above transformation law for $\Lambda$ takes on the following form
$\Lambda'(X')=\Lambda(X)$ and $\Lambda$ is a scalar field. 
In the identical way as for $b^{k}$ we
show that $\Lambda=\Lambda(\partial_{i}\partial_{j}\phi)$.
So, $\Lambda$ is one of the Kronecker's invariants of the matrix 
$(\partial_{a}\partial_{b}\phi)$.
Now, we come back to the equation and easily show that it can be covariant 
if and only if $k$ is a constant.
We get, then, the the Schr\"odinger equation which after the standard 
notation of constants has the form
\begin{equation}\label{ec2} 
\Big[ \frac{\hslash^{2}}{2m}\delta^{ij}\partial_{i}\partial_{j} 
+i\hslash\partial_{t} -m\phi+\Lambda\Big]\psi = 0,
\end{equation}
with the $\theta$ in $T_{r}$ given by
\begin{displaymath}
\theta=\frac{m}{2\hslash} \int_{0}^{t} \dot{\vec{A}}^{2}(\tau) \, {\ud{\tau}} 
+\frac{m}{\hslash}\dot{A}_{i}x^{i}.
\end{displaymath}   
 Note that the the inertial mass
$m$ in the equation is equal to the parameter at the gravitational potential.
That is, the gravitational mass must be equal to the inertial mass.

It is remarkable that the wave equation would be covariant with respect 
to the Galilean group even when $m_{i} \neq m_{g}$ because the potential 
$\phi$ transforms as a scalar under this group. This is a consequence 
of the Galilean covariance of the ordinary Schr\"odinger equation with 
a scalar potential. 

In the paper \cite{AHJW} the same wave equation has been derived with 
the additional result $\Lambda = 0$ as a consequence of the equivalence 
principle applied to the wave equation. 

Equality of inertial and gravitational 
masses in the quantum regime was verified
experimentally, see \cite{COW}, \cite{Kasevich}.
In the beautiful experiment of Kasevich and Chu \cite{Kasevich} 
the E\"otv\"os parameter for the sodium atom  was estimated to be 
$\leq 10^{-6}$.

\section{Comparison with a classical point particle}

The same result can be obtained for the classical particle, but under slightly
stronger assumptions. The second theorem of N\"other connected with the Milne 
covariance and with the gauge freedom of the potential gives us the identities 
equivalent to the equation of motion of the particle, that is, the geodetic 
equation. Because many authors think after Einstein and Infeld 
that the particle equations of motion  has to be separately and \emph{a priori}
postulated in the Newton-Cartan theory, we present the 
argument in details. They thought that
the equations of motion of a 
particle are independent of the space-time geometry structure, but that 
the equations follows from the gravitational field equations.
Then, they argue that the Poisson equation is linear and by this cannot contain
the equations of motion for matter, e.g. for particles. However, the
equations of motion of matter are not connected with the gravitational
field equations itself, but first of all with the space-time structure and
the covariance conditions, that is, the form-invariance for the Lagrange 
density function of matter.  The knowledge of the gravitational field
equations is also needed  when the particles are not testing.
This is true in general, in the non-relativistic as well 
as in  the relativistic theory of gravity. 
Now we present this in details.
Suppose we have a matter fields $b^{A} = b^{A}(X)$ with the 
Lagrange density function $\mathcal{L}= \mathcal{L}(X, b^{A},\phi)$.
We consider the N\"other identities. In our case the covariance transformations
depend on arbitrary functions of the time only, but not of all space-time 
coordinates. Compare the Milne transformations or the time dependent gauge 
transformation of the potential $\phi \to \phi + \dot{\epsilon}(t)$.
This arbitrariness does not allows us to obtain the differential identities,
but it is sufficient to obtain some integral identities, if we suppose
that the matter density goes to zero sufficiently fast when the space 
coordinates goes to infinity. Namely, let us denote the arbitrary functions
of time defining the covariance transformations by 
$\epsilon^{i}= \epsilon^{i}(t)$.
The variations of the fields $ \equiv \{b^{A}, \phi\}$ 
under the action of the transformation are as follows
\begin{displaymath}
\overline{\delta}y^{\mathcal{A}} = c^{\mathcal{A}}_{i}\epsilon^{i} + 
d^{\mathcal{A}\mu}_{i}\partial_{\mu}\epsilon^{i} + 
g^{\mathcal{A}\mu \nu}_{i}\partial_{\mu}\partial_{\nu}\epsilon^{i}.
\end{displaymath} 
Let us write $\mathcal{L_{A}} \equiv\{\mathcal{L}_{A},\mathcal{L}_{0}\}$ 
for the Euler-Lagrange derivatives of $\mathcal{L}$ with respect
to $y^{\mathcal{A}}$, $\mathcal{L}_{0}$ is the Euler-Lagrange 
derivative with respect to $\phi$.
Then we have the following N\"other integral identities:
\begin{displaymath}
\int_{R^{3}}\Big[\mathcal{L_{A}}c^{\mathcal{A}}_{i}
-\partial_{\mu}(\mathcal{L_{A}}d^{\mathcal{A}\mu}_{i}) 
+\partial_{\mu}\partial_{\nu}(\mathcal{L_{A}}g^{\mathcal{A}\mu \nu}_{i})\Big]
\, \ud^{3}x \equiv 0. 
\end{displaymath}
The identities for the gauge transformation $\phi \to \phi + \dot{\epsilon}(t)$
and the Milne transformations $x^{i} \to x^{i} + \epsilon^{i}(t)$ read 
\begin{equation}\label{gauge}
\frac{d}{dt}\Big[\int_{R^{3}}\mathcal{L}_{0} \, \ud^{3}x\Big] \equiv 0, 
\end{equation}
\begin{equation}\label{milne1}
\frac{d^{2}}{dt^{2}}\Big[\int_{R^{3}}(-\mathcal{L}_{0}x_{i}) \, \ud^{3}x\Big]
\equiv \int_{R^{3}}\mathcal{L}_{0}\partial_{i}\phi \, \ud^{3}x, 
\end{equation}
by virtue of $\mathcal{L}_{A}=0$. At this place we have to make additional 
assumption as compared to quantum level, that
in the limit for the point particle moving along a trajectory 
$z_{i} = z_{i}(t)$ we have
\begin{equation}\label{classical}
\mathcal{L} = \mathfrak{L}(x_{i},t)\delta(x_{i}-z_{i}),
\end{equation}
where $\delta$ is the three-dimensional Dirac delta function.
Suppose first that the matter is minimally coupled: $\mathcal{L}$
does not depend on derivatives of the potential.
From the identity (\ref{gauge}) it follows that 
$\mathcal{L}_{0} = const \, \delta(x-z) \equiv -m\delta(x-z)$.
Inserting this to the identity (\ref{milne1}) one obtains
\begin{equation}\label{geodetic}
m\ddot{z}_{i} = -m\partial_{i}\phi.
\end{equation} 
Now, suppose that the matter is not minimally coupled and the Lagrange density
does depend on the first degree derivatives of the potential. We assume, 
however, that the Lagrange density does not depend on the second and the 
higher derivatives of the potential. As a consequence of this assumption and 
from the identity (\ref{gauge}) we get 
$\mathcal{L}_{0} = -const \,\delta(x-z)- \partial_{i}(Q^{i}\delta(x-z))$, where
$Q^{i} = \partial\mathfrak{L}/\partial(\partial_{i}\phi)$. The identity 
(\ref{milne}) gives
\begin{equation}\label{general}
m\ddot{z}_{i} = -m\partial_{i}\phi + Q^{j}\partial_{i}\partial_{j}\phi
+ \ddot{Q}_{i}.
\end{equation} 
As is seen from this equation the quantity 
$W_{i} = Q^{j}\partial_{i}\partial_{j}\phi+\ddot{Q}_{i}$ has to be a vector. 
In accord to our general assumption $W_{i}$ is an algebraic function of 
$\phi$, its derivatives up to the third order, $\dot{z}_{i}$ and $\ddot{z}_{i}$.
An analysis similar to that presented above shows that 
$W_{i} = W_{i}(\partial_{k}\phi+\ddot{z}_{i}, 
\partial_{t}\partial_{j}\partial_{k}\phi
+\dot{z}^{a}\partial_{a}\partial_{j}\partial_{k}\phi, 
\partial_{a}\partial_{b}\phi,\partial_{a}\partial_{b}\partial_{c}\phi)$.
Taking this into account and the specific form of the $W_{i}$ 
in (\ref{general}) we prove that $W_{i}=0$ and the equation (\ref{general}) 
takes on the form of the geodetic equation (\ref{geodetic}). 
We have not analyzed the situation in which the Lagrange density $\mathcal{L}$
of matter can \emph{a priori} depend on second order derivatives of $\phi$.
Because it is natural to assume that the field equations of matter and gravity
are of second order at most it is natural to assume that $\mathcal{L}$ does
not depend on second and higher order derivatives of the potential. 

So, if one assume that the Lagrange density does not depend on second and 
higher order derivatives of $\phi$, and that the equations of matter are 
generally covariant and can be constructed in a local way from the geometry 
of space-time, then the equations of motion for the particle are determined
by the space-time geometry. Moreover, the equations are in accord with 
the equivalence principle. 
Summing up, we have shown exactly the same for the quantum particle\footnote{The 
equality of both masses can be obtained of course by considering the WKB 
approximation of the corresponding classical particle. But this is the whole 
point: in the classical equations of motion mass cancels out on both sides 
if the gravitational mass is equal to the inertial one. WKB approximation
is just equivalent to classical mechanics. In full Schr\"odinger equation
the mass does not cancel. The identities (\ref{gauge}) 
and (\ref{milne1})  are also true for the
Lagrange density function $\mathcal{L}$  of the Schr\"odinger equation. 
The assumption (\ref{classical})
can be fulfilled approximately for a finite time interval. So, we get
the equivalence principle at the classical level without
imposing any condition on the inertial mass in the Schr\"odinger equation,
because the identities (\ref{gauge}) and (\ref{milne1}) do not involve
the kinetic term which contains the inertial mass. So, there is no immediate
relation between the the equality of the inertial and the gravitational masses
for a  classical particle and the equality for a quantum 
particle. Therefore the ability to prove the equality is nontrivial.}. 
Note, that the equation (\ref{ec2}) cannot be derived 
from the Lagrange density which does not contain the 
second derivatives of the field $\phi$ if $\Lambda \ne 0$. 
Our result is nontrivial if one takes into account (1) the 
observation of Trautman \cite{Tr} that the equivalence principle can be 
violated by a field with the Lagrange function containing
the first degree derivatives of $\phi$ and (2) the equation
of motion for any matter field cannot be derived in this way.  

\vspace{1cm}

The author is indebted for helpful discussions to A. 
Staruszkiewicz, A. Herdegen, M. Je\.zabek and W. Kopczy\'nski. The paper 
was supported by the Polish State Committee for Scientific 
Research (KBN) grant no. 5 P03B 093 20.


\begin{thebibliography}{99}

\vspace{.5cm}

\bibitem{Car} \'E. Cartan, Ann.\'Ec. Norm. Sup {\bf 40}, 755 (1923); 
{\sl ibid.} {\bf 41}, 1 (1924).
\bibitem{Tra} A. Trautman, Comptes Rendus Acad. Sci. Paris {\bf 247}, 617 
(1963).
\bibitem{Dau} G. Da\u utcourt, Acta Phys. Polon. {\bf B21}, 755 (1990).
\bibitem{And} J. L. Anderson, {\sl Principles of Relativity Physics}, AP, N-Y, 
London (1967), see chap.4; M. Friedman, {\sl Foundations of Space-Time 
Theories}, Princeton Univ. Press, Princeton (1983).
\bibitem{AHJW} A. Herdegen and J. Wawrzycki, Phys. Rev. {\bf D 66},
044007, (2002).
\bibitem{Bievre} S. De Bi\'evre, Class. Quantum Grav {\bf 6}, 731, (1989). 
\bibitem{Milne} A. Milne, Q. J. Math. {\bf 5}, 64, (1934); C. Duval, Class.
Quantum Grav. {\bf 10}, 2217, (1993); J. Christian, Phys. Rev. {\bf D 56}, 
4844, (1997). 
\bibitem{Bar} V. Bargmann, Ann. Math. {\bf 59}, 1, (1954).
\bibitem{JW} J. Wawrzycki, math-ph/0301005.
\bibitem{COW} R. Colella, A. N. Overhauser and S. A. Werner, Phys. Rev. Lett. 
{\bf 34}, 1472, (1975).
\bibitem{Kasevich} M. Kasevich and S. Chu, Phys. Rev. Lett. {\bf 67}, 181, 
(1991).
\bibitem{Nesvi} V. V. Nesvizhevsky \emph{et al.}, Nature {\bf 415}, 297, 
(2002). 
\bibitem{DK} C. Duval, H. P. K\"unzle, Gen. Rel. Grav. {\bf 16}, 333 (1984).
\bibitem{Kuch} K. Kucha\v r, Phys. Rev. {\bf D 22}, 1285, (1980).
\bibitem{Waw} J. Wawrzycki, Int. Jour. of Theor. Phys. {\bf 40}, 1595 (2001).
\bibitem{Acz} J. A. Schouten, {\sl Tensor analysis for physicists}, Oxford 
(1951), Oxford Univ. Press.
\bibitem{Tr} A. Trautman, In: {\sl Gravitation: an introduction to current 
research}, edited by L. Witten, (John Wiley $\&$ Sons, Inc., New 
York, London 1962), p. 169. Compare especially the comments placed on 
the page 174. 


\end{thebibliography}
\end{document}